\documentclass{article}[11]

\usepackage{graphics,latexsym}
\usepackage{graphicx}
\usepackage{amsmath, amssymb,amsthm}

\begin{document}

\title{\bf On the Near-field of an Antenna \\
\bf and the Development of New Devices}
\author{Daniele Funaro}

\date{}
\maketitle

\centerline{Department of Mathematics} \centerline{University of
Modena and Reggio Emilia}
\centerline{Via Campi 213/B, 41125
Modena (Italy)} \centerline{daniele.funaro@unimore.it}

\begin{abstract}
Dipole antennas have been invented a hundred years ago.
Nevertheless, what really happens in their proximity during
emission (the so-called near-field) is still an open question and
the many explanations put forth are not fully convincing. Subject
to specific conditions the signal present on the conductor assumes
the properties of a pure electromagnetic wave. We would like to
give our point of view on the modality of this transition, with
the hope of providing suggestions for ameliorating or projecting
new devices.
\end{abstract}

\vspace{.8cm}
\noindent{Keywords: dipole antennas, near-field, non solenoidal fields.}
\par\smallskip

\noindent{PACS:  41.20.Jb, 84.40.Ba, 02.30.Jr}

\renewcommand{\theequation}{\thesection.\arabic{equation}}

\par\medskip
\setcounter{equation}{0}
\section{Non divergenceless fields}


We start by introducing the model equations proposed in
\cite{funarol}. They include Maxwell's equation as a particular
case allowing an impressive enlargement of the space of solutions.
The main important consequence of this approach is the possibility
of handling photons (i.e., non-diffusive electromagnetic emissions
having compact support) in the classical fashion, reopening the
old diatribe on the wave-particle duality. This is not however the
subject that we are discussing here. We want instead to analyze
the impact of this alternative formulation on the study of simple
antennas.
\par\smallskip

Let us first remark that the revised model shares with Maxwell's
all the good properties and characterizations, presenting no weak
points in this comparison. One of the crucial steps is to assume
that, even in vacuum, the electric field may have divergence
different from zero. The reasons for this choice are thoroughly
documented in \cite{funarol} and successive papers (see
\cite{funaro3}, \cite{funaro1}, \cite{funaro4}). We would like to
stress here that, also in the framework of simple applications,
this assumption is not just a theoretical exercise, but a real
necessity. We invite then the reader to accept possible
explanations regarding the functioning mechanism of antennas, in
the light of a solid theory where non divergenceless (i.e.: non
solenoidal) electric fields are allowed.
\par\smallskip

Let us begin by writing down the equations in the case of
{\sl free-waves}, which form a subset of all possible electromagnetic
manifestations. Denoting by ${\bf E}=(E_1,E_2,E_3)$  the electric
field and by ${\bf B}=(B_1,B_2,B_3)$ the magnetic field, we have:
\begin{equation}\label{eq:sfem1}
\frac{\partial {\bf E}}{\partial t}~=~ c^2{\rm curl} {\bf B}~
-~\rho {\bf V}
\end{equation}
\begin{equation}\label{eq:sfbm1}
\frac{\partial {\bf B}}{\partial t}~=~ -{\rm curl} {\bf E}
\end{equation}
\begin{equation}\label{eq:sfdb1}
{\rm div}{\bf B} ~=~0
\end{equation}
\begin{equation}\label{eq:slor1}
\rho ~({\bf E}+{\bf V}\times {\bf B}) ~=~0
\end{equation}
where we set: $\rho ={\rm div}{\bf E}$ (this is a definition, not
a ruling equation). Moreover,  $c$ denotes the speed of light and
${\bf V}=(V_1,V_2,V_3)$ is a velocity field in such a way that the
triplet $({\bf E},{\bf B},{\bf V})$ is right-handed. The magnitude
of ${\bf V}$ is equal to $c$. Equation (\ref{eq:sfem1}) turns out
to be the Amp\`ere law for a free flowing immaterial current with
density $\rho$. It is clear that by imposing $\rho =0$ (this now
becomes an equation) we reobtain the classical Maxwell's system.
Equations (\ref{eq:sfbm1}) and (\ref{eq:sfdb1}) do not need
explanation. Equation (\ref{eq:slor1}) characterizes free-waves
and, as we will see later, can be assimilated to a sort of
geometrical constraint. More explanations regarding its physical
meaning can be found in \cite{funarol} and \cite{funaro3}.
\par\smallskip

We do not waste time here for a discussion of the physical and
theoretical properties of the above set of equations. We just
mention a few meaningful facts. A continuity equation for $\rho$
is already contained in equation (\ref{eq:sfem1}). In fact, by
taking its divergence, one obtains:
\begin{equation}\label{eq:cont}
\frac{\partial \rho}{\partial t}~=~ -{\rm div} ({\rho \bf V})~=~
-\rho ~{\rm div}{\bf V}~-~\nabla \rho\cdot {\bf V}
\end{equation}
By scalar multiplication of (\ref{eq:slor1}) by ${\bf V}$ and
${\bf B}$, we get respectively:
\begin{equation}\label{eq:ort}
{\bf E}\cdot{\bf V}~=~0 ~~~~~~~~~~~~~~{\bf E}\cdot{\bf B}~=~0
\end{equation}
Afterwards, with the same calculations relative to the Maxwell's
case, the following energy relation holds:
\begin{equation}\label{eq:ener}
\frac{1}{2} \frac{\partial}{\partial t} (\vert {\bf E}\vert^2 +
c^2 \vert {\bf B}\vert^2)+ c^2{\rm div}({\bf E} \times {\bf
B})~=~0
\end{equation}
where ${\bf E}\times {\bf B}$ is the Poynting vector.
\par\smallskip

In the system of time-space coordinates where $x_0=ct$, we can
consider the electromagnetic stress tensor $T^{\alpha \beta}$. It
is known that Maxwell's equations are compatible with the
conservation law: $\partial_\alpha T^{\alpha \beta}=0$. As a
matter of fact, by summing up on the index $\beta$, one has:
\begin{equation}\label{eq:ten1}
\frac{\partial T^{0\beta}}{\partial x_\beta}=\frac{1}{2c}
\frac{\partial}{\partial t}
(\vert {\bf E}\vert^2 + c^2 \vert {\bf B}\vert^2)+ c~{\rm div}({\bf E}
\times {\bf B})
\end{equation}

$$\left(\frac{\partial T^{1\beta}}{\partial x_\beta},
\frac{\partial T^{2\beta}}{\partial x_\beta},
\frac{\partial T^{2\beta}}{\partial x_\beta}\right)=
\left(\frac{\partial {\bf B}}{\partial t}+{\rm curl}{\bf E}\right)\times{\bf E}$$
\begin{equation}\label{eq:ten2}
-\left(\frac{\partial {\bf B}}{\partial t}-c^2{\rm curl}{\bf B}\right)\times{\bf B}
~+~{\bf E}~{\rm div}{\bf E}~+~c^2{\bf B}~{\rm div}{\bf B}
\end{equation}
Therefore, if Maxwell's equations in vacuum are satisfied all the
above expressions are zero (recall (\ref{eq:ener})). On the other
hand, by adding and subtracting the term $({\rm div}{\bf E}){\bf
V}$, (\ref{eq:ten2}) takes the form:
$$\left(\frac{\partial T^{1\beta}}{\partial x_\beta},
\frac{\partial T^{2\beta}}{\partial x_\beta},
\frac{\partial T^{2\beta}}{\partial x_\beta}\right)=
\left(\frac{\partial {\bf B}}{\partial t}+{\rm curl}{\bf E}\right)\times{\bf E}$$
\begin{equation}\label{eq:ten3}
-\left(\frac{\partial {\bf B}}{\partial t}-c^2{\rm curl}{\bf B}
+({\rm div}{\bf E}){\bf V}\right)\times{\bf B}
+({\bf E}+{\bf V}\times {\bf B}){\rm div}{\bf E}~+~c^2{\bf B}~{\rm div}{\bf B}
\end{equation}
which is now compatible with the new set of equations, so
justifying their origin.
\par\smallskip

All these considerations can be extended to general metric spaces
by writing the equations in covariant form. Lorentz invariance and
the existence of a Lagrangian have been also considered, though
these issues are out of the scope of this paper.
\par\smallskip

Finally, let us note that a vector wave equation for ${\bf E}$ is
not available. We claim that this is not a trouble but a key point
in the success of the new approach. Solutions of the wave equation
tend to diffuse all around (without energy dissipation anyway) and
this is one of the reasons for the impossibility to include
compact-support solitary waves in a classical theory. According to
the new set of equations, we can differentiate (\ref{eq:sfem1})
with respect to time. Using (\ref{eq:sfbm1}) one gets:
$$
\frac{\partial^2 {\bf E}}{\partial t^2}~=~-c^2{\rm curl}({\rm
curl} {\bf E})~-~\frac{\partial (\rho {\bf V})}{\partial t}
$$
\begin{equation}\label{eq:ondev}
=~c^2\Delta {\bf E}~-~c^2\nabla \rho~-~\frac{\partial \rho
}{\partial t}{\bf V}-~\rho \frac{\partial {\bf V} }{\partial t}
\end{equation}
From the continuity equation (\ref{eq:cont}), one obtains:
\begin{equation}\label{eq:ondev2}
\frac{\partial^2 {\bf E}}{\partial t^2}~ =~c^2\Delta {\bf
E}~+~[(\nabla \rho \cdot {\bf V}){\bf V}~-~c^2\nabla \rho]~+~\rho
{\bf V}~{\rm div}{\bf V}-~\rho \frac{\partial {\bf V} }{\partial t}
\end{equation}
yielding the usual wave equation for $\rho =0$. In order to
understand the importance of (\ref{eq:ondev2}) let us consider a
simple example in cartesian coordinates. Suppose that ${\bf
E}=(u,0,0)$, for a certain $u$ not depending on $y$. Suppose that
${\bf V}=(0,0,c)$ (note that ${\bf E}$ is orthogonal to ${\bf V}$
in accordance to (\ref{eq:ort})). We have: ${\rm div}{\bf V}=0$,
$\partial {\bf V}/\partial t =0$ and $\rho =\partial u/\partial
x$. By substituting into (\ref{eq:ondev2}) we get:
\begin{equation}\label{eq:ondes}
\frac{\partial^2 u}{\partial t^2}~ =~c^2~\frac{\partial^2
u}{\partial z^2}
\end{equation}
This is a scalar wave equation where the solution is shifting
along the $z$-axis, which is the one individuated by the vector
${\bf V}$. This is different from the canonical wave equation,
that should also include a second derivative with respect to the
variable $x$. The term $[(\nabla \rho \cdot {\bf V}){\bf
V}-c^2\nabla \rho]$ on the right-hand side of (\ref{eq:ondev2}) is
exactly placed with the purpose of cancelling from the Laplacian
the unwanted derivatives. A similar result for $\rho$ is obtained
by plugging ${\bf V}=(0,0,c)$ in (\ref{eq:cont}), which brings to:
$\partial \rho /\partial t =-c\partial \rho /\partial z$, where
the derivative with respect to $x$ does not appear.
\par\smallskip

Hence, as far as the variable $x$ is concerned, we have no
restrictions on $u$. Let us also observe that the only solution
compatible with $\rho =0$, i.e. belonging to the solution space of
Maxwell's equations in vacuum, imposes that $u$ must be constant
in the variable $x$. We get in this way an unbounded plane wave,
showing how little are the degrees of freedom when working with
fields with zero divergence.

\section{The classical dipole antenna}

Let us now face the discussion of the classical dipole antenna.
The equations introduced in the previous section teach us how to
get perfect spherical wave-fronts with no approximation. These are
good for the description of the travelling wave in the far-field.
\par\smallskip

As customary, computations are carried out in spherical
coordinates $(r,\theta ,\phi)$. We define the fields:
\begin{equation}\label{eq:esf}
{\bf E}~=~\Big( 0, ~\frac{cf(\theta )}{r}g(t-r/c), ~0\Big)~~~~~~
{\bf B}~=~\Big( 0,~ 0, ~-\frac{f(\theta )}{r}g(t-r/c)\Big)
\end{equation}
where $f$ and $g$ are arbitrary functions. The profile $g$
controls the time-dependent amplitude of the wave and $f$ its
energy distribution on each spherical wave-front. In addition, we
set ${\bf V}=(c,0,0)$ (the transport velocity is along the radial
direction). By direct substitution in the model equations, one
checks that this setting is an admissible solution.
\par\smallskip

We have that $({\bf E}, {\bf B}, {\bf V})$ is an orthogonal
triplet with:
\begin{equation}\label{eq:rect}
\vert {\bf E}\vert ~=~\vert c{\bf B}\vert
\end{equation}
compatibly with the impedance of free space of about $376\Omega$.
\par\smallskip

With the setting in (\ref{eq:esf}), the only way to get a solution 
satisfying $\rho =0$ is to impose
$f(\theta )=1/\sin\theta$, causing a singularity along the
vertical axis. Of course, if we drop the condition $\rho =0$, we
enjoy much more freedom. In particular, one can set $f(\theta
)=\sin\theta$, which is the standard far-field solution of a half
wave-length dipole. One could argue that the last solution is
obtained from the Maxwell's model by eliminating negligible terms
(take the Hertz solution for an infinitesimal dipole and get rid
of the components decaying faster than $1/r$ for $r\rightarrow +\infty$). From
our point of view, why insisting on approximated solutions when
the new model accepts them as exact solutions?
\par\smallskip

The real problem is to understand what happens in the near-field
and, evidently, our dipole cannot be infinitesimal. The emitting
device is going to be the segment $[-d/2, d/2]$ in the direction
of the vertical axis. For simplicity we assume that $2d$ is the
wave-length of an emitted monochromatic frequency $\omega /2\pi$.
In other words, we suppose that $g(t-r/c)=\sin \omega (t-r/c)$ and
$d=\pi c/\omega$.
\par\smallskip

The first problem is to discuss boundary conditions. The antenna
can be seen as an oscillating chord, with standing nodes at the
endpoints. From the solution of a one-dimensional wave equation,
one recovers that the time-dependent datum to be assigned on the
vertical segment is: $\sin (\omega t)\cos (\omega r/c)$, for
$r\leq d/2$. This behavior is in general referred to the current
intensity circulating within the wire.
\par\smallskip

We would like to follow a different path. To this end we take into
consideration the model equations. As observed at the end of the
previous section, the electric field comes from the resolution of
a wave equation where on the right-hand side there is not a full
Laplacian in spherical coordinates but only second derivatives in
the direction of motion. We note that ${\bf E}$ does not depend on
$\phi$ and that the only component different from zero is the one
relative to the angle $\theta$. Thus,  ${\bf E}$ is of the form
$(0,u,0)$, where $u$ restricted to the antenna satisfies a
``truncated'' wave equation of the following type:
\begin{equation}\label{eq:waver}
\frac{\partial^2 u}{\partial
t^2}~=~\frac{c^2}{r^2}\frac{\partial}{\partial r}\left(
r^2\frac{\partial u}{\partial r}\right)
\end{equation}
The solution of the above equation is related to spherical
Bessel functions. In explicit terms, we find out that:
\begin{equation}\label{eq:solu}
u~=~\sin (\omega t)\frac{\cos (\omega r/c)}{r}~~~~~~{\rm for}~~
r\leq d/2
\end{equation}
This is true up to a multiplicative constant which has to be
regulated according to the power of the feeding source. The
proposed behavior is slightly different from the usual one since
there is an extra $r$ at the denominator, producing a faster decay
as the electric signal moves towards the extremes of the antenna.
On the other hand, such an alternative proposal takes into account
that the device is not separated from the general context, being
embedded in a spherical environment. In (\ref{eq:solu}), we have a
singularity for $r$ tending to zero; we know however that there is
a small gap between the two arms of the antenna, so that one can
safely stay away from the critical region.
\par\smallskip

If $u$ is the boundary datum, we may ask ourselves what is going
to be the behavior of the fields as we leave the antenna. We make
the following guess:
\begin{equation}\label{eq:esfs}
{\bf E}~=~\Big( 0, ~c\frac{\sin (\omega t)\cos(\omega
r/c)}{r\sin\theta}, ~0\Big)~~~~~~ {\bf B}~=~\Big( 0,~ 0, ~-\frac{\cos
(\omega t)\sin(\omega r/c)}{r\sin\theta}\Big)
\end{equation}
This is in perfect agreement with Maxwell's equations (i.e.: $\rho
=0$). To avoid singularities, we have to assume that $r\sin\theta$
has to be bigger that a quantity $\epsilon$, which is linked to
the width of the antenna wire. Moreover, we shall require that
$r\leq d/2$. This means that we are staying in a sort of spherical
resonant cavity. The electric and magnetic fields have a phase
difference of $90$ degrees, therefore this cannot be a travelling
wave (note instead that the fields in (\ref{eq:esf}) have no phase
lag). This oscillating situation is associated with the boundary
constraints and the corresponding fields do not go anywhere.
Indeed the boundary datum does not ``shift'', representing on the
contrary a standing wave.
\par\smallskip

\begin{center}
\begin{figure}[h]
\centerline{\hspace{-.3cm}\includegraphics[width=16.cm,height=12.cm]{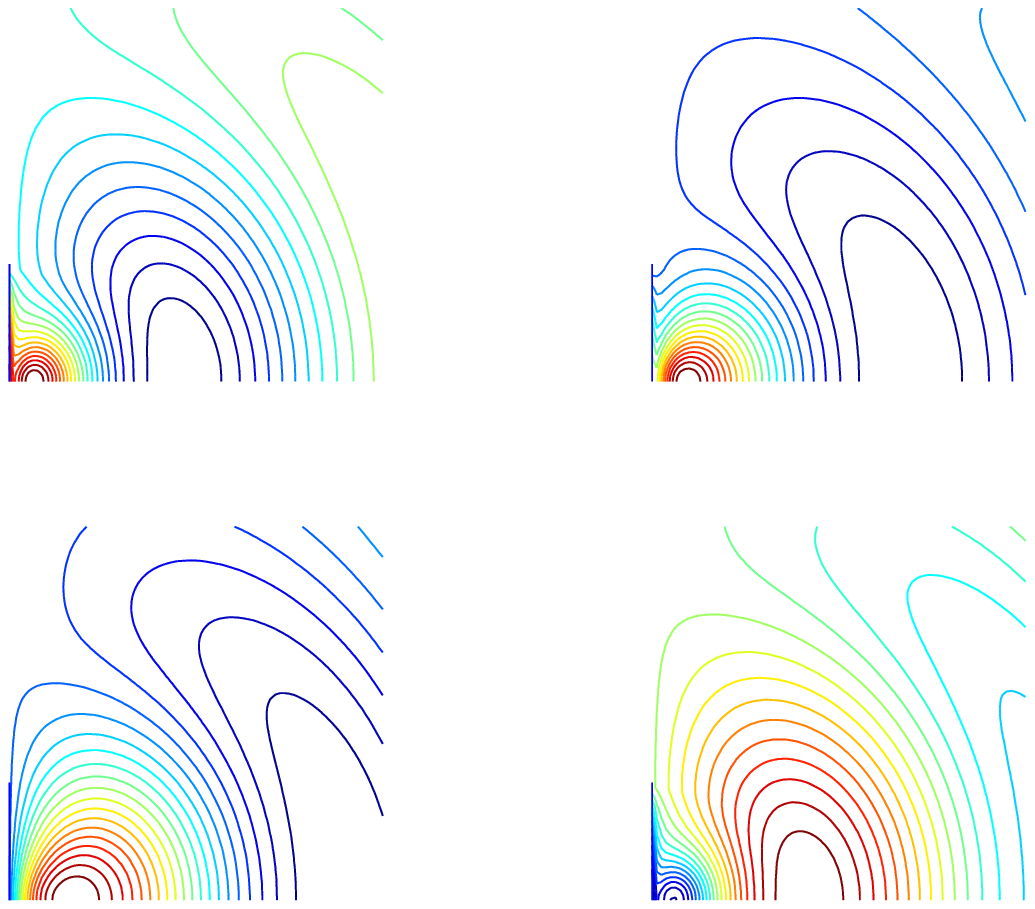}}
\vspace{-.5cm}
\begin{caption}{\small Magnitude of $E_2$ in (\ref{eq:sef}) at different
times, for a suitable choice of the constants $k_1$ and $k_2$. The
upper arm of the dipole is visible on the bottom-left. There is a
kind of fictitious bubble separating the guided region from the
one where the wave is free. The reader should not confuse the
above lines, which are the level sets of the scalar function
$\vert {\bf E}\vert$, with the lines of force of the field, which
are distributed on perfect spherical surfaces (${\bf E}$ only has
the $\theta$ component) as seen in figure 2.}
\end{caption}
\end{figure}
\end{center}
\vspace{-.5cm}

We can pose at this point the crucial question. Why and how the
fields circulating in the neighborhood of the antenna start
becoming a free electromagnetic emission? The answer we propose is
based on a linear combination of the fields in (\ref{eq:esf})
and (\ref{eq:esfs}). Let us observe that
the operation of summing up a solution having $\rho =0$ and a
solution having $\rho \not =0$ is compatible with the model
equations, also if these are of nonlinear type. Explicitly, for
the second component of the electric field we have:
\begin{equation}\label{eq:sef}
E_2~=~k_1 \frac{f(\theta )}{r}\sin \omega (t-r/c)~+~k_2\frac{\sin
(\omega t)\cos(\omega r/c)}{r\sin\theta}
\end{equation}
where $k_1$ and $k_2$ are arbitrary constants to be set up based
on the power transferred to the antenna device and its geometrical
properties (width of the wires and size of the central gap).
For the half wave-length dipole we suggest to take $f(\theta )=\sin \theta$.
The evolution of this kind of wave can be seen in figure 1.
Figure 2 shows the orientation of ${\bf E}$.
\par\smallskip

To recap, the first part in (\ref{eq:sef}) is the effective
travelling wave and continues to survive for $r>d/2$. Its boundary
conditions are zero and remain zero also when there is no more
boundary, in order to guarantee the continuity of the spherical
wave at the poles. To be more precise, the function $f$ in
(\ref{eq:sef}) should vanish for $\epsilon$ and $\pi -\epsilon$, where
$2\epsilon$ is the diameter of the conducting wire.
The second part is active only for $r<d/2$ and
is compatible with the boundary conditions. It is the standard
field produced by an oscillating current in a wire under
resonant conditions and follows
Maxwell's equations. The global solution is continuous but not
smooth for $r=d/2$. The real wave probably follows a better
behavior, which also depends on the degree of smoothness of the
wires at the endpoints.
This is what we can do using simple analytic
functions; the only way to get more reliable information is to use
numerical computations.
\par\smallskip

Let us add more explanations. For reasons that will be clarified
later, in the narrow gap of the feeding area
electric and magnetic fields are generated. Part of this message
displays no phase lag,  becoming the actual
signal that in the end propagates as a free-wave. The rest of the effective energy will
be dissipated along the arms. Understanding what happens in this
gap is crucial for the comprehension of the entire mechanism of
emission. We return on this subject in section 4.
\par\smallskip

From the production site to the open space,
there is the need of a transition zone (the near-field) before the
wave can effectively be expelled. As a matter of fact, the
guided-wave flowing in the feed line (usually a coaxial wire) satisfies Maxwell's
equations ($\rho =0$). This is possible because one can define the
internal boundary conditions accordingly. But, when the wave is
completely free there is no boundary and we need to assume that
$\rho\not =0$, entering the regime of the modified model
equations. We reinforce this statement by recalling that one
cannot comb a sphere (the magnetic field follows the parallels
maintaining zero divergence, the electric field is tangent to the
meridians and is obliged to be non divergenceless). We need then
an appropriate zone to realize smoothly the passage from $\rho =0$
to $\rho\not =0$, and spread the non divergenceless property along
spherical surfaces.
\par\smallskip

\begin{center}
\begin{figure}[h]
\centerline{\includegraphics[width=12.cm,height=8.7cm]{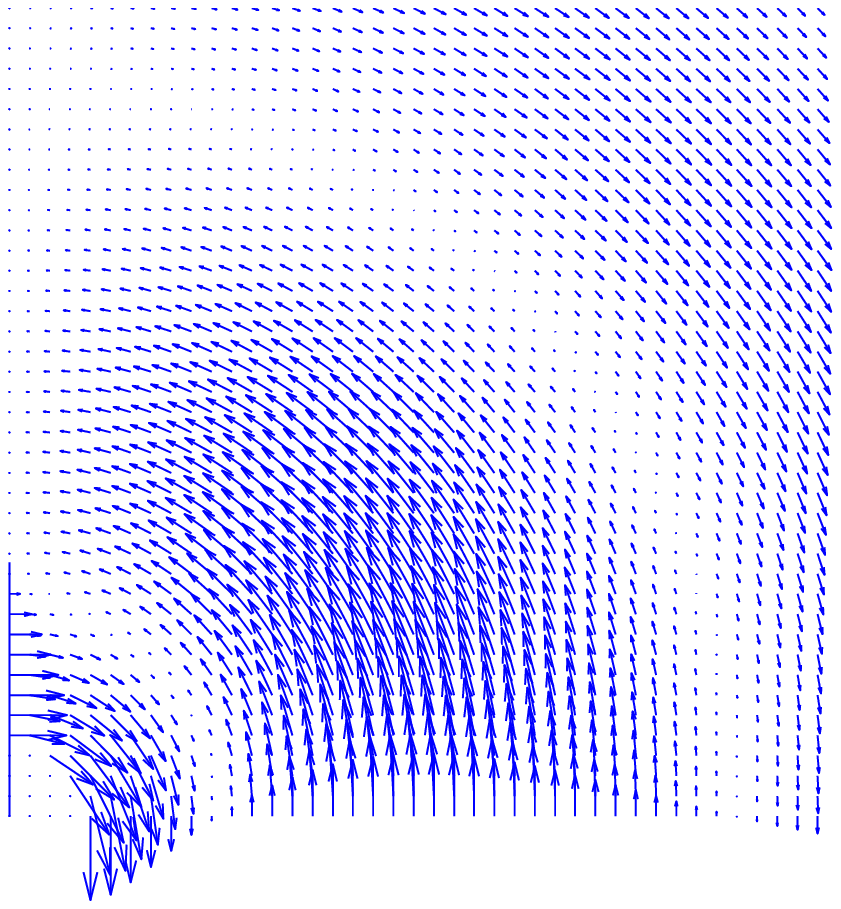}}
\vspace{-.3cm}
\begin{caption}{\small Orientation of the electric field in the case
of the second picture of figure 1. The magnetic field is
orthogonal to the page. To facilitate the view, a portion near the
origin has been omitted since the intensity there is to strong.
The signal is always zero when it reaches the north pole, and will
stay zero in all the successive free evolution. The Poynting
vector is always directed outbound. }
\end{caption}
\end{figure}
\end{center}
\vspace{-.5cm}

The arms of the antenna are used as guides to delimitate the area.
Note however that they are not  ``wave-guides'' as it is usually
intended, because the information on these conductors, being
related to a standing wave, does not shift. Such a distinction is
not emphasized in the literature, where antenna devices are often
treated as a prolongation of the wave-guide carrying the original
signal (see for instance \cite{schelk}, chapter XI). This
simplifies a lot the analysis especially in numerical
computations, but does not explain the mechanism of the antenna,
based essentially on its resonant properties with consequent local
production of standing waves. What we are examining here is
exactly the possibility to combine both effects.
\par\smallskip

Due to properties that we will try to better clarify later in section 4, an
electric field tends to ``stick'' at the boundary surface. This
adherence is very pronounced, at the level that it is very
difficult for a conductor to get rid of the signal. This should
explain why the transition area has to be a resonant cavity with a
fictitious spherical boundary supplied through the antenna device.
Indeed, let us note that the second term on the right-hand side of
(\ref{eq:sef}) vanishes for $r=d/2$, i.e. all the points of the
corresponding spherical surface are of nodal type. The electric normal
component adheres to the antenna wires, but fades when approaching the
endpoints attenuating the strength of the link (see figure 2). As
the endpoints are reached ($r=d/2$) the boundary field is zero.
The electromagnetic wave has no more contact with the emitting
device and, since it has been prepared to support fields with
nonzero divergence, can finally fly away.
\par\smallskip

Note that the Poynting vector is always oriented radially; the
energy flow moves from the central point at the speed of light and
does not change trajectory. This conclusion is different from that
obtainable theoretically by examining divergenceless fields (see
for instance \cite{jackson}), where the radiation patterns in the
near-field tend to approach the arms of the antenna. Nevertheless,
the whole procedure might not be as simple as we are claiming, so
that more remarks will be added in the coming sections.

\section{Coaxial antennas}

We now work in cylindrical coordinates $(r,\phi ,z )$. It turns
out that a family of solutions of the whole set of model equations
is:
$${\bf E}~=~\Big(cf(r)g(t-z/c),~0, ~0\Big)$$
\begin{equation}\label{eq:sol2}
~~{\bf B}~=~\Big(0,~f(r)g(t-z/c), ~0\Big)~~~~~~~~~~~{\bf
V}~=~(0,~0, ~c)
\end{equation}
where $f$ and $g$ can be arbitrary. As the orientation of ${\bf
V}$ testifies, the wave shifts along the $z$-axis solving equation
(\ref{eq:ondes}). In general, the electric field is non
divergenceless and we can only obtain $\rho =0$ if $f(r)=1/r$.
Inside a coaxial cable we can handle the situation $\rho =0$ by
assuming suitable boundary conditions. In this case we are
supposing that there is no dielectric between the internal wire
and the external shield. The lines of force of the magnetic field
are coaxial circumferences.
\par\smallskip

As the signal reaches the end of the cable one should expect an
ejection like that of water from a garden hose. It is well-known
however that this does not happen. Why? Because we are
disregarding the two principles stated in the previous section.
First of all, if we want the wave to get out and survive it is
necessary to create a transition area in order to ensure the
condition $\rho \not =0$. In fact, as the effect of the boundary
constraints ceases, the electric field has to damp transversally
to zero and this cannot be realized if $\rho$ remains equal to
zero. The second issue concerns the contact with the boundary. The
strength of the link must diminish until it vanishes. The wave
must be accompanied to the exit through the creation of a resonant
cavity. Part of the cavity boundary must be fictitious,
representing the open way out of the ejected wave.
\par\smallskip

In order to better quantify the behavior at the end of a coaxial
cable, a further extension of the model equations is needed (see
\cite{funarol} and \cite{funaro3}). Equation (\ref{eq:slor1}) is
improved as follows:
\begin{equation}\label{eq:slorm}
\rho \left(\frac{D{\bf V}}{Dt}~+~\mu({\bf E}+{\bf V}\times {\bf B})\right)
~=~-\nabla p
\end{equation}
which is the non-viscous Euler equation for the velocity field
${\bf V}$. Among the unknowns we have now a new entry, i.e. the
potential $p$, that, up to multiplicative dimensional constant,
can be regarded as a pressure (the same radiation pressure
measured when light hits an obstacle). The substantial derivative
$D {\bf V}/Dt$ is defined as $\partial {\bf V} /\partial t +({\bf
V}\cdot \nabla ){\bf V}$. We also have a coupling constant $\mu$,
dimensionally equivalent to charge/mass, which has been estimated
to be approximately equal to $2.3\times 10^{11}$ Coulomb/Kg.
Note the strong analogy between (\ref{eq:cont}), (\ref{eq:slorm}) and
the equations ruling plasma physics (see \cite{jacksonl}, p.491).
The novelty here is that there are no particles involved, but
everything is related to the evolution of pure fields. The
density of particles is now substituted by $\rho$.
\par\smallskip

Equation (\ref{eq:slorm}) amounts to an abstract generalization of
the Lorentz law. When the field balance ${\bf E}+{\bf V}\times
{\bf B}=0$ is not realized, there are accelerations imparted to
the flow-field ${\bf V}$, which may change both in magnitude and
direction. The corresponding wave does not freely propagate and
the energy follows new velocity patterns. Condition ${\bf E}\cdot
{\bf V}$ (see (\ref{eq:ort})) is not fulfilled, as it happens for
instance during the propagation in a medium different from vacuum.
In this case the electric field has a component in the direction of motion.
We have free-waves when $D{\bf V}/Dt=0$ and $p=0$; in this case,
if no external factors are acting on the wave, the signal
propagates along straight rays.
\par\smallskip

Let us try to understand what happens as a signal arrives at the
extreme of a coaxial cable. At the very end, a discontinuity tends
to be created in the outgoing transversal electric field, due to
the sudden disappearance of the boundary. This has the effect of
breaking the balance ${\bf E}+{\bf V}\times {\bf B}=0$, creating a
region where $\rho \not=0$. Thanks to (\ref{eq:slorm}) there is a
deviation of the velocity stream-lines with a corresponding
generation of pressure. It can be shown (using the same arguments
followed in \cite{funarol}, p.117, also confirmed by the numerical
tests in \cite{funaro1}) that the divergence of ${\bf V}$
(initially zero in the coax) becomes positive inducing a spread of
the signal. The electric field rotates and sticks on the outer
conducting shield (note that the magnetic lines of force remain
circumferences). This forms a kind of bottleneck. Part of the
energy is rejected and returns back to the source. Therefore, no
significant electromagnetic emission is observed. Estimates of the
radiated power, obtained by using classical arguments, are given in
\cite{mcdonald2}.
\par\smallskip

Thus, an antenna is needed to drive the signal in the proper way
(similar arguments are developed in \cite{mcdonald}, but always
within the context of fields with zero divergence). Let us also
observe that, by virtue of equations (\ref{eq:slorm}), a finer
analysis can be made on the dipole. There, we assumed that the
stream-lines  determined by ${\bf V}$ were straight and
originating from the small gap between the wires, but this
interpretation is probably too simplistic. Due to the presence of
a dielectric inside the feeding cable, condition (\ref{eq:rect})
is not realized. This means that in the near-field the
transformation of the signal has to take care of the impedance
adaptation. When the correct ratio of the intensities of the
electric and magnetic fields is not respected, one has ${\bf
E}+{\bf V}\times {\bf B} \not=0$. As we said, this implies that an
acceleration is imparted on the velocity field modifying a bit the
trajectories. Through a feedback process, an adjustment of the
electromagnetic  field is made via (\ref{eq:sfem1}) and
(\ref{eq:sfbm1}) until the right impedance is reached.
The behavior of photons in proximity of matter has been investigated
in \cite{keller}, section 3.6, where estimates based on truncated expansions
point out the difficulties emerging when one imposes the condition 
$\rho =0$ in the vacuum regions around matter.
\par\smallskip

One also has to deal with the fact that the arms are not perfect
conductors and in their neighborhood the speed of propagation
of the signal is less than $c$.
Moreover, because of the asymmetry of the term ${\bf E}+{\bf
V}\times {\bf B}$, the acceleration vector may change orientation
according to the sign of the electric field, which varies in time,
recalling the sprouts of an oscillating sprinkler. In other terms,
${\bf V}$ is not going to be a stationary velocity field. Although
the far-field tends asymptotically to be distributed on spherical
surfaces, the energy can slightly oscillate transversally with the
frequency of the emitted wave. In order to recover more
information, numerical simulations can be carried out by solving
the full set of equations involving the simultaneous computation
of ${\bf E}$, ${\bf B}$, ${\bf V}$ and $p$. In this way, we are
not solving a wave equation, but a set of equations that couple
electromagnetism with non-viscous fluid dynamics, where the
velocity field indicates the direction of the energy flow. These
are: (\ref{eq:sfem1}), (\ref{eq:sfbm1}), (\ref{eq:sfdb1}) and
(\ref{eq:slorm}), forming a closed system with a total of 10
relations and 10 unknowns.
\par\smallskip

Going back to the coaxial case, let us see if we can devise a
system to actually expel the signal. The simplest idea is to use a
device as shown in figure 3, where the hot wire is connected to a
vertical wire (a quarter wave-length long, i.e.: $d/2=\pi
c/2\omega$) and the shield to a horizontal conductive disk. For a
given frequency, the setting has to form a kind of resonant
hemisphere in such a way that the intensity of the electric field,
at the end of the wire and at the border of the disk, is zero. The
electromagnetic signal coming out from a small gap between the
wire and the disk is expected to be driven along conical pathways.
\par\smallskip

In order to compute the optimal size of the disk, one has to solve
a wave equation on it. We assume that the electric field stays
orthogonal to the disk and we denote by $u$ the only component
different from zero. We then eliminate from the Laplace operator
in cylindrical coordinates the derivatives contributions with
respect to $z$ and $\phi$. Of course, this is just an
approximation since we still do not know the exact shape of the
cavity. In truth, the resonant zone should be determined through a
global process where also the vertical wire is taken into
consideration. Nevertheless, we arrive at the equation:
\begin{equation}\label{eq:wavec}
\frac{\partial^2 u}{\partial
t^2}~=~\frac{c^2}{r}\frac{\partial}{\partial r}\left(
r~\frac{\partial u}{\partial r}\right)
\end{equation}
whose solutions are related to Bessel functions. Our suggestion is
to take $u=\sin (\omega t) Y_0(\omega r/c)$, where $Y_0$ is a
Bessel function of the second kind (presenting a singularity at
the origin). The radius of the disk has to be determined in
correspondence of the first zero of $Y_0$, which is $0.893
c/\omega$ (we recall the $2\pi c/\omega$ is the total wavelength).
Hence, the ratio between the length of the wire and the radius of
the disk is about $1.76$. Of course, these conclusions are referred
to the hypothetical case of a perfect device.
\par\smallskip

\begin{center}
\begin{figure}[h]\vspace{.5cm}
\centerline{\includegraphics[width=8.cm,height=8.cm]{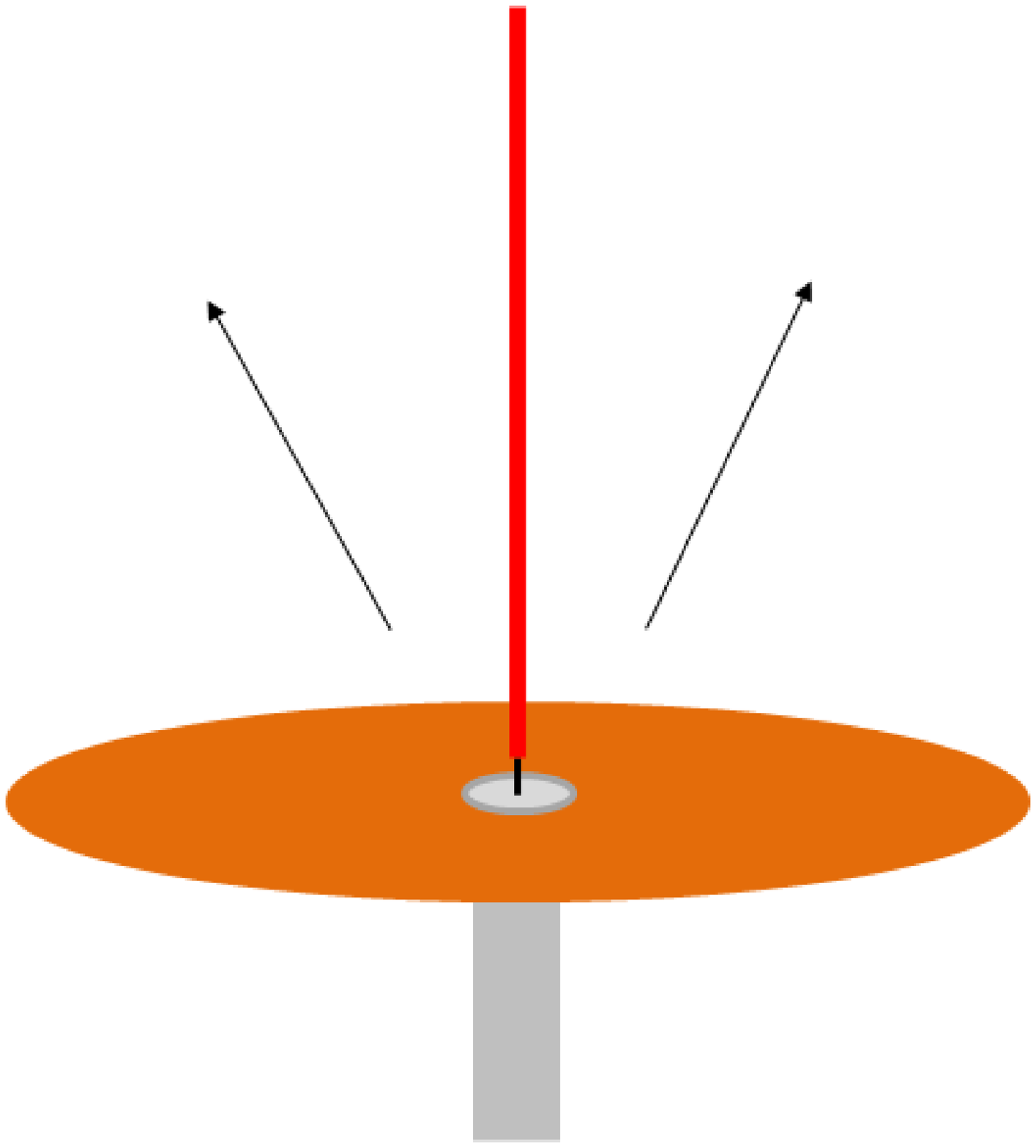}}
\begin{caption}{\small The output of a coaxial cable is connected
to a plate and a vertical wire (monopole antenna). The ejected signal is expected to
follow conical paths. The magnetic field is divergenceless and
distributed on coaxial circumferences. For a given frequency, the
optimal power emanated should be obtained when the whole system
forms a resonant cavity, having both the boundary of the disk and
the endpoint of the wire as nodal sets.}
\end{caption}
\end{figure}
\end{center}
\vspace{-.2cm}

The above information may be used for dimensioning the ground of a
mono\-pole antenna (for an overview the
reader is addressed for instance to \cite{weiner}).
Early experimental results, for many choices of the disk diameter,
were conducted in \cite{maley}. From the
theoretical viewpoint, the monopole is usually regarded as a half-space
dipole. Reliable theoretical estimates are thus available
(see also \cite{johnson}, p.4-28, in particular for what concerns the
radiation patterns).
The fact that the radiation is directed into the upper half tends
to be justified by the reflecting properties of the ground.
Our explanation is indeed quite different from the qualitative viewpoint. 
Take however into account that the new set of equations includes
the space of solutions of the standard Maxwell's system, therefore
all previous results and conclusions still hold true.


\section{Further details and speculations}

We recall once again that requiring the divergence of the electric
field to be nonzero is not just a mathematical abstraction but a
hypothesis one cannot avoid. If we mechanically move the plates of
a charged parallel capacitor (no dielectric in between), the
intensity of the field varies from a constant to another. The process
does not occur instantaneously, since the information needs some time to
propagate. Inevitably, there will be regions inside the capacitor
where the electric field, whose lines of force are assumed to be
orthogonal to the plates, is not constant and $\rho\not =0$. It is
true that $\rho$ is going to be extremely small due to the fact
that the information travels at the speed of light. But, when we
deal with a coaxial cable operating at high frequency, a change of
size, perhaps due to a joint, may have an appreciable influence on
$\rho$.
\par\smallskip

The study of a capacitor under the influence of oscillating
fields is clearly explained with plenty of details in the
Feynman's book (see \cite{feynman}, vol.2, chapter 23).
The antenna gap between the two arms of the dipole forms a tiny capacitor,
which is exposed to an oscillating potential, whose frequency is
not the resonant one but not too far from it. The presence of values of  $\rho$
different from zero is not negligible at this stage. A similar condition
is presumably encountered in the
sparks generated by an Hertzian coil, also present in the early
versions of the Marconi's antenna.
In \cite{feynman}, for a capacitor composed by two disks connected at a generator
with a pair of wires, the analysis is carried out as a function
of the distance from the symmetry axis. However, no indications
about the longitudinal behavior are provided, so skipping the
conclusions about $\rho$.
\par\smallskip

For an infinitely extended parallel capacitor, subject to a homogeneous
difference of potential $V_0\sin (\omega t)$ we can
explicitly show some computations. The horizontal plates are located
at $z=\pm a$. It is the matter of solving the wave equation
$\partial^2 V/\partial t^2= c^2\partial^2 V/\partial z^2$,
with the boundary conditions $V(t,\pm a)=\pm{\textstyle{\frac{1}{2}}}
V_0\sin (\omega t)$. We then get the solution:
\begin{equation}\label{eq:solcon}
V(t,z)~=~\frac{V_0 \sin (\omega t)\sin (\omega z/c)}{2\sin (\omega a/c)}
\end{equation}
that clearly also depends on $z$.
By taking the gradient one recovers the third component of the electric field that, up
to multiplicative constant, behaves as $E_3=\sin (\omega t)\cos (\omega z/c)$.
The divergence of this field is certainly nonzero. If the argument
of the cosinus is small (plates at short distance or low-frequency) the
electric field  only changes with time, remaining practically constant
in space. Instead, in the gap of a dipole antenna (which is in general not
so small compared to the entire device) $\rho$ starts assuming
meaningful values.
\par\smallskip

By computing the curl of the field ${\bf E}=(0,0,E_3)$, by (\ref{eq:sfbm1})
one recovers that the induced magnetic field ${\bf B}$ must satisfy
$\partial {\bf B}/\partial t =0$. The whole setting is incompatible
with Maxwell's equations, even if both ${\bf E}$ and ${\bf B}$ solve
independently a vector wave equation. Here the inconsistency is due
to the assumption that the infinite plates are uniformly supplied.
So, as done in \cite{feynman}, we should suppose that the plates are disks
supplied at the center. Computations are now more involved, since together
with the dependence on the variable $z$ we also have to allow
${\bf E}$ to be non-parallel. Having $\rho\not =0$, we cannot rely on
Maxwell's equation and we are obliged to use the new set of equations.
The signal produced in this semi-resonant regime is
neither fish nor fowl, being a combination of some fields in phase with
other fields out of phase.
The results turn out to be more complex, but at the same time more realistic
(so we think).
What happens just outside the small gap of the antenna is difficult
to predict without numerical simulations. As we said, part of
the signal flows along the arms and another part is ejected, after
passing through a transition area. A correct design of the feeding gap
is a necessary step to guarantee the optimality of the output; in fact
this small component has a crucial role in the antenna performances
(see for instance \cite{collin} or \cite{johnson}).
\par\smallskip


In section 2, we mentioned about the ``sticking'' properties of the electric
field on a metallic surface. Let us formalize better this concept.
Consider for simplicity an electric field ${\bf E}$ orthogonal to a surface
and not depending on time
(for instance the constant field inside a parallel capacitor
in the stationary regime). The field is zero inside the conductor, implying
the existence of a discontinuity. For a more accurate analysis, one
should enter the atomic structure and explain the reasons of
this sharp variation. We stay however vague with this respect.
We just assume that, within a distance comparable to a few
Angstroms, the electric field drops from a constant to zero,
producing a sharp boundary layer. Of course $\rho\not =0$
within this neighborhood.  Take now ${\bf B}=0$ and ${\bf V}=0$
in the model equations. The only relation surviving is (\ref{eq:slorm}),
which yields: $\mu\rho {\bf E}=-\nabla p$. Using that ${\bf E}$
is orthogonal to the surface, one gets: $p=-\mu \vert {\bf E}\vert^2$.
Therefore,  a sort of negative ``pressure'' can be felt near the surface, also
if there are no physical particles exerting it.
We claim that this is responsible of the effective mechanical
movement of the charged plates, commonly attributed to Coulomb attraction,
explaining how such an action-at-a-distance may occur in practice.
As far as the antenna is concerned, this reasoning should clarify
why the signal remains glued to the device until ${\bf E}=0$.
These heuristic arguments are very unconventional but efficacious.
\par\smallskip

The last issue we would like to discuss here is what happens at the
end of the antenna arms. We said that each endpoint is a node for the
transversal electric field. This cannot be true for the magnetic
field which is out of phase of 90 degrees. One may notice that
this situation is in contrast with the absence of current at
the endpoints. But, if one argues in terms of pure electromagnetic
fields, thus excluding the existence of an actual flow of particle
charges (such as electrons), there is no contradiction. We only
have to explain what happens to the magnetic component as the
wire ends. We recall that our wire has a diameter equal to
$2\epsilon$. We guess that on top of the upper arm the nonzero oscillating magnetic field
is coupled with a newborn electric field, both enclosed in a small bubble
forming a thin cap of the same diameter of the wire. Both
fields decay fast to zero with the the distance from the endpoint.
Such isolated region does not interfere with the emitted wave.
Again, these speculative considerations can be verified with
appropriate numerical tests.
One can also try to figure out the  behavior of the fields
in the farthest sections, when the length of the dipole is not
a multiple of half wave-length (see for instance \cite{johnson}, p.4-6).

\section{Future projects}

In this section we would like to fantasize about some ambitious
projects. Let $f$ in (\ref{eq:sol2}) be a continuous function
vanishing outside a specified interval. Thus, the support of the
corresponding wave is a cylinder. We are now dealing with a
perfect monochromatic electromagnetic soliton. Can a signal of
this type be produced in nature? Can we build a source capable of
emitting such a wave?  Do antennas with infinite directivity
exist? If this was true, we would be able to communicate
point-to-point without reaching other targets and with no energy
dissipation.
\par\smallskip

According to the experience this seems not feasible. Maybe it is
true: the question has no answer. Maybe it is only matter of
exploiting new territories. Now we have a new set of model
equations. Now we also have a new set of recipes teaching us how
to design an antenna. The result must come from a symbiosis of
laboratory tests and numerical computations. The agenda is open.
The creation of very directive signals, certainly useful in
applications, would also validate the theoretical model.
\par\smallskip

The suggestions given here are addressed to optimize the antenna
configuration for a given emitted frequency. Of course we expect
that the same apparatus may continue to work within a certain
range of frequencies (bandwidth), though with some losses.
Ordinary applications show us that antennas can do a reasonable
job under conditions that are far from optimality. Nevertheless,
there is always an inverse relation to be observed, connecting
frequency and the reciprocal of the size of the emitting device.
It is difficult to get around this restriction. 
\par\smallskip

It seems
impossible for instance to produce radio-waves on the range of one
KHz with a tool a few centimeters wide. There are several reasons
indicating that this is hard to achieve. We first need a source where
oscillating electric and magnetic fields are produced with no
phase lag. This cannot be a capacitor under the action of an alternate voltage.
The effect would be very mild (see previous section).
However, we can independently combine
electric and magnetic fields, generated by individual sources, to
get the desired effect. So, this problem can be somehow
circumvented. Afterwards, there is the question of pre-regulating
$\rho$ in order to prepare the wave to the open space. As we said,
a capacitor subject to an alternate field must produce internal
regions with $\rho\not =0$, although $\rho$ is terrifically small
at low frequencies. Interposed materials with variable dielectric
constants may help to this purpose. Anyway, the most difficult task
is to get rid of the signal and in order make it possible one cannot account on
resonance phenomena. We know however what should be done: the
electric signal has to reach the emission site with zero intensity
on the edge of the device, in order to lose contact with it. Once
discovered the ill, somebody may come out with a medicine. Solving
the problem of transmitting low-frequency radio-waves in small
space is important to allow the transfer of energy at a distance
and in large amount. The top is to  realize this goal with the
maximum directivity.

\end{document}